\documentstyle[preprint,epsf]{ptptex}
\title{
\hfill
\parbox{5cm}{\normalsize 
\normalsize UT-841\\
}\\
Critical Exponents of O(N) Scalar Model at  Temperatures below the
Critical Value using Auxiliary Mass Method 
}

\author{
Kenzo {\sc Ogure}\footnote{E-mail address: ogure@icrr.u-tokyo.ac.jp} 
and Joe {\sc Sato}$^{*,}$\footnote{E-mail address:
 joe@hep-th.phys.s.u-tokyo.ac.jp}
}

\inst{
Institute for Cosmic Ray Research, University of Tokyo, 
Tanashi, Tokyo 188-8502, Japan
\\
$^*$Department of Physics, School of Science, University of Tokyo,
Tokyo 113, Japan}

\recdate{
\today
}

\abst{
We investigate a phase transition of the O(N) invariant scalar model using
the auxiliary mass method. We determine the critical exponent $\beta$ by
calculating an effective potential below the critical temperature. This
work follows that of a previous paper.\cite{OJ2}  }

\begin{document}
\maketitle
Phase transitions at finite temperature are important phenomena in
particle physics, cosmology and condensed matter physics.  For
example,  the QGP phase is produced in heavy ion
collisions.\cite{QGP} Some phase transitions occured in
the early universe.\cite{KT} The electro-weak
phase transition in particular plays an important role in the electro-weak
baryogenesis scenario\cite{Kux2} and gives some constraints to models
of elementary particle physics.\cite{CKN,RS,CQW} \ We also see a great
number of phase transitions in condensed matter physics. In the
present paper, we investigate an O(N) invariant scalar model which
corresponds to many condensed matter systems, for example alloys,
superfluids, and binary liquids.\cite{ZJ}

To investigate such phase transitions, we can use finite
temperature field theory, which is based only on a statistical
principle.  However, we often have an infrared divergence and cannot
obtain reliable results using perturbation theory at finite
temperature.\cite{Fen}  To overcome this problem, we used the
auxiliary-mass method,\cite{DHL,IOS,OJ1} and calculated an effective
potential and critical exponents of the O(N) invariant scalar model above
the critical temperature $T_c$
in a previous paper.\cite{OJ2} \ We did not investigate
at a temperature below $T_c$  for two
reasons, numerical instability and the lack of
computer power.  In this work we have overcome these problems, and
we calculate an effective potential and critical exponents of the O(N)
invariant scalar model below the critical temperature.

We explain the idea of the auxililary-mass method.  Since we can
calculate a reliable effective potential for temperatures $T\ll
\frac{m}{\lambda}$ using perturbation theory,\cite{Fen} first we
assume the mass as $m\sim T$ and calculate an effective potential.
This potential is reliable if the coupling constant, $\lambda$, is
small.  We next extrapolate the effective potential to that of a true
mass using a non-perturbative evolution equation.  Finally, we
determine the necessary physical quantities. We determine critical
exponents below $T_c$ in the present paper.

Applying this method to the O(N) invariant scalar model, the
Euclidean Lagrangian density is given by
\begin{eqnarray}
     {\cal L}_{E}=-\frac{1}{2}
     \left(\frac{\partial \phi_a}{\partial \tau}\right)^{2}
     -\frac{1}{2}(\mbox{\boldmath $\nabla$} \phi_a)^{2}
     -\frac{1}{2}m^{2}\phi_a^{2}
     -\frac{\lambda}{4!}(\phi_a^{2})^{2}
     +J_a\phi_a + c.t.\ 
    \label{lag}
\end{eqnarray}
Here, the $J_a$ are external source functions, and the index $a$ runs
from 1 to N. We assume that the coupling constant $\lambda$ is small, 
and therefore the perturbation theory at zero temperature is reliable.
We first calculate the effective potential at an auxiliary large mass
$m=M\sim T$ at the one-loop level as
\begin{eqnarray}
     V&=&\frac{1}{2}M^{2}\bar{\phi}^{2}
     +\frac{\lambda}{4!}\bar{\phi}^{4}
     +\frac{T}{2\pi^{2}}
     \int^{\infty}_{0}dr\ r^{2}\log \left[
     1-\exp\left(-\frac{1}{T}\sqrt{r^{2}+M^{2}+
     \frac{\lambda}{2}\bar\phi^{2}}\right)\right]\nonumber \\
     &&+(N-1)\ \frac{T}{2\pi^{2}}
     \int^{\infty}_{0}dr\ r^{2}\log \left[
     1-\exp\left(-\frac{1}{T}\sqrt{r^{2}+M^{2}+
     \frac{\lambda}{6}\bar\phi^{2}}\right)\right].
    \label{ini}
\end{eqnarray}
Here $\bar\phi$ is a field expectation value.  We leave only
the finite-temperature part of the equation because we can ignore the
zero temperature part due to the small coupling constant.  We note
that the daisy-resummation is not necessary because of the large mass.
We then construct a non-perturbative evolution equation which connects
the effective potential at an auxiliary large mass, $m^2\sim T^2$, and
that of the true mass, $m^2=-\mu^2$.  Since we have constructed this for the
O(N) invariant scalar model in a previous work,\cite{OJ2} we present
only the result:
\begin{eqnarray}
     \frac{\partial V}{\partial m^{2}}&=&
     \frac{1}{2}\bar{\phi}^{2}+\frac{1}{4\pi^{2}}
     \int^{\infty}_{0}dr\ r^{2}\frac{1}{\displaystyle \sqrt{r^{2}
     +\frac{\partial^{2}V}{\partial\bar\phi^{2}}}}
     \frac{1}{\displaystyle \exp\left(\frac{1}{T}\sqrt{r^{2}
     +\frac{\partial^{2}V}{\partial\bar\phi^{2}}}\right)-1}\nonumber\\
     &&+\frac{N-1}{4\pi^{2}}
     \int^{\infty}_{0}dr\ r^{2}\frac{1}{\displaystyle \sqrt{r^{2}
     +\frac{1}{\bar\phi}\frac{\partial V}{\partial\bar\phi}}}
     \frac{1}{\displaystyle \exp\left(\frac{1}{T}\sqrt{r^{2}
     +\frac{1}{\bar\phi}
     \frac{\partial V}{\partial\bar\phi}}\right)-1}.
    \label{evo}
\end{eqnarray}
This partial differential equation is solved with the initial
conditions (\ref{ini}) numerically.\\

\begin{figure}
\unitlength=1cm
\begin{picture}(16,8)
\unitlength=1mm
\centerline{
\epsfxsize=10cm
\epsfbox{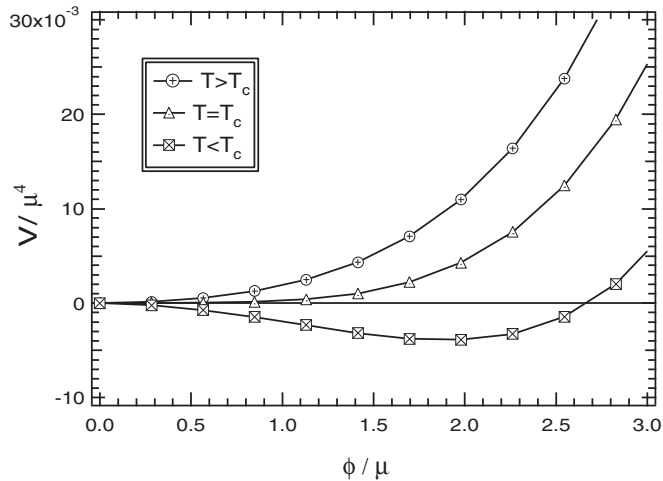} 
} 
\end{picture}
\caption{The effective potential obtained by the auxiliary-mass
  method ($N=4,\lambda =0.01$). A second-order phase transition occurs at
  the critical temperature. Similar behaviour is observed at 
  other values of N and $\lambda$.}
\label{cripot}
\end{figure}

We display the effective potential for N=4 around $T_c$ in Fig. 
\ref{cripot}, and we find that the phase transition is of second
order.  The same behaviour is found for other values of N.  This is
consistent with other analyses using lattice field theory and
renormalization group theory.\cite{ZJ} We find that the auxiliary-mass
method satisfactorily deals with the problem of the infrared divergence.

\begin{figure}
\unitlength=1cm
\begin{picture}(16,8)
\unitlength=1mm
\centerline{
\epsfxsize=10cm
\epsfbox{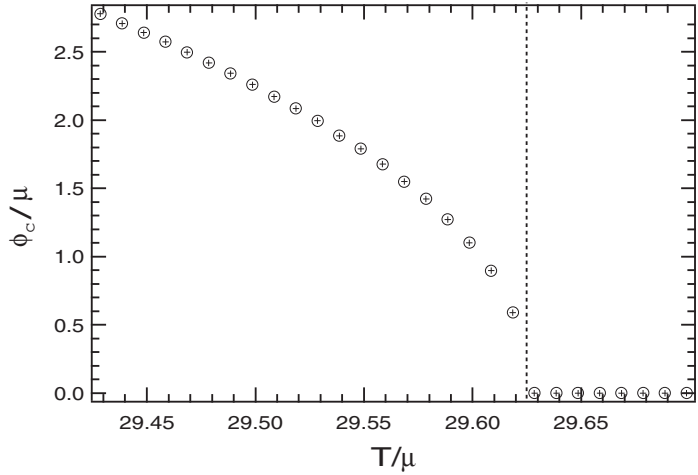} 
} 
\end{picture}
\caption{Stable field expectation value $\phi_c$ as a function of the 
temperature $T$ ($\lambda=0.01$). $\phi_c$ decreases monotonically and vanishes
smoothly as the temperature increases.  }
\label{xmin1}
\end{figure}

\begin{figure}
\unitlength=1cm
\begin{picture}(16,8)
\unitlength=1mm
\centerline{
\epsfxsize=10cm
\epsfbox{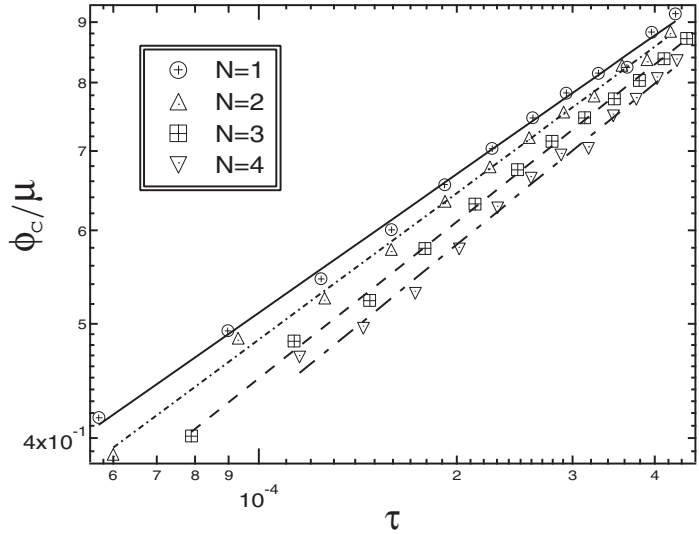} 
} 
\end{picture}
\caption{Log-scale plots of $\phi_c-\tau$ ($\lambda=0.01$). The
data were fit by linear functions with gradients corresponding to $\beta$
for each N. We find that the exponent $\beta$ is larger for larger N.}
\label{logxmin}
\end{figure}

  We next determine the
critical exponents and study how well the auxiliary-mass method works.
Since we have investigated this model above
$T_c$ previously, obtainig the critical exponents
$\gamma$ and $\delta$,\cite{OJ2} we investigate below 
$T_c$ and determine the critical exponent $\beta$ here.
The critical exponent, $\beta$ relates an order parameter,
$\phi_c$, to a reduced temperature, $\tau\equiv\frac{T_c-T}{T_c}$, as 
\begin{eqnarray}
     \phi_c \propto \tau^{\beta}.
    \label{beta}
\end{eqnarray}
The order parameter $\phi_c$ as a function of reduced
temperature $\tau$ is presented in Fig.\ref{xmin1} for N=4. Similar
behaviour for other values of N is found.
Since the order parameter $\phi_c$
vanishes smoothly at $T_c$, we find that the phase
transition is of second order.  We next plot $\phi_c$ as a
function of $\tau$ in Fig.\ref{logxmin} for various N. These data
appear linear with different gradients, corresponding 
to $\beta$ for each N. The exponent, $\beta$ is
larger for larger values of N.

We summarise the results of a present paper and a previous
paper\cite{OJ2} in Table.\ref{com}. The values of $\beta, \gamma$ and  
$\delta$ are much better than the Landau approximation
and the dependence on N is close 
to the most reliable  value (MRV). There are, however, slight differences
between our results and the MRV, which are caused by an
approximation  in deriving  Eq.\ref{evo}.\footnote
{ An improvement of the approximation is underway.  }

In conclusion, we have investigated the O(N) invariant scalar model
using the auxiliary-mass method and have obtained good results both
qualitatively and quantitatively.  These results suggest that the
auxiliary-mass method is an effective tool at finite temperature.  We
were able to investigate not only second order phase transitions but
also first order phase transitions since the finite-temperature field
theory is based only on a statistical principle.  We therefore believe
that this is one of the most powerful methods to investigate a weak
first order phase transition and models which have end-points : cubic
anisotropy, the abelian Higgs model and the standard model.\footnote
{ We are preparing to apply this method to these model presently.  }

\begin{table}[ht]
\caption{The critical exponents $\beta$, $\gamma$ and $\delta$ obtained in
a present paper and the previous paper.  Those of Landau approximation
(LA) and the most reliable values (MRV) are also summarised. We used
the results of the six-loop approximation using the Pad\'{e}-Borel
resummation for the MRV here.} {\small
\begin{center}
\begin{tabular}{lccc}
\hline\hline
      & $\beta$ (LA,MRV)  & $\gamma$ (LA,MRV)   & $\delta$ (LA,MRV) \\
\hline
N$=$1 \cite{Ant}&0.39 (0.5, 0.327)  & 1.37  (1, 1.239)  &   4.0  (3, 4.8) \\
N$=$2 \cite{Ant}& 0.41 (0.5, 0.348) &1.47   (1, 1.315)  &   4.2   (3, 4.8)\\
N$=$3 \cite{Ant}& 0.44 (0.5, 0.366) &1.60    (1, 1.386) &   4.4  (3, 4.8) \\
N$=$4 \cite{Ant} & 0.45 (0.5, 0.382) &1.66    (1, 1.449) &  4.4  (3, 4.8)  \\
\hline
\end{tabular}\\
\end{center}
}
\label{com}
\end{table}

The authors are supported by JSPS fellowship.

\end{document}